\documentclass[bibyear]{aa}  

\usepackage{graphicx}
\usepackage[dvipsnames]{xcolor}
\usepackage{marvosym}
\usepackage{natbib}
\bibpunct{(}{)}{;}{a}{}{,} 
\usepackage{xspace}
\usepackage{enumitem}
\usepackage[colorlinks=true, linkcolor=blue, citecolor=blue, urlcolor=blue]{hyperref}
\usepackage{subcaption,booktabs}
\usepackage{comment}
\usepackage[thinc]{esdiff}
\usepackage{soul}
\usepackage{txfonts}
\usepackage{pifont}
\usepackage{multirow}
\usepackage{makecell}
\usepackage{float}

\defcitealias{ishiyama2025}{IH25}
\defcitealias{hartwig2022}{H22}
\defcitealias{sana2012}{S12}
\defcitealias{sb2013}{SB13}

\usepackage{txfonts}
\usepackage{pifont}

\newcommand{\orcidicon}[1]{\href{https://orcid.org/#1}{\includegraphics[width=11pt]{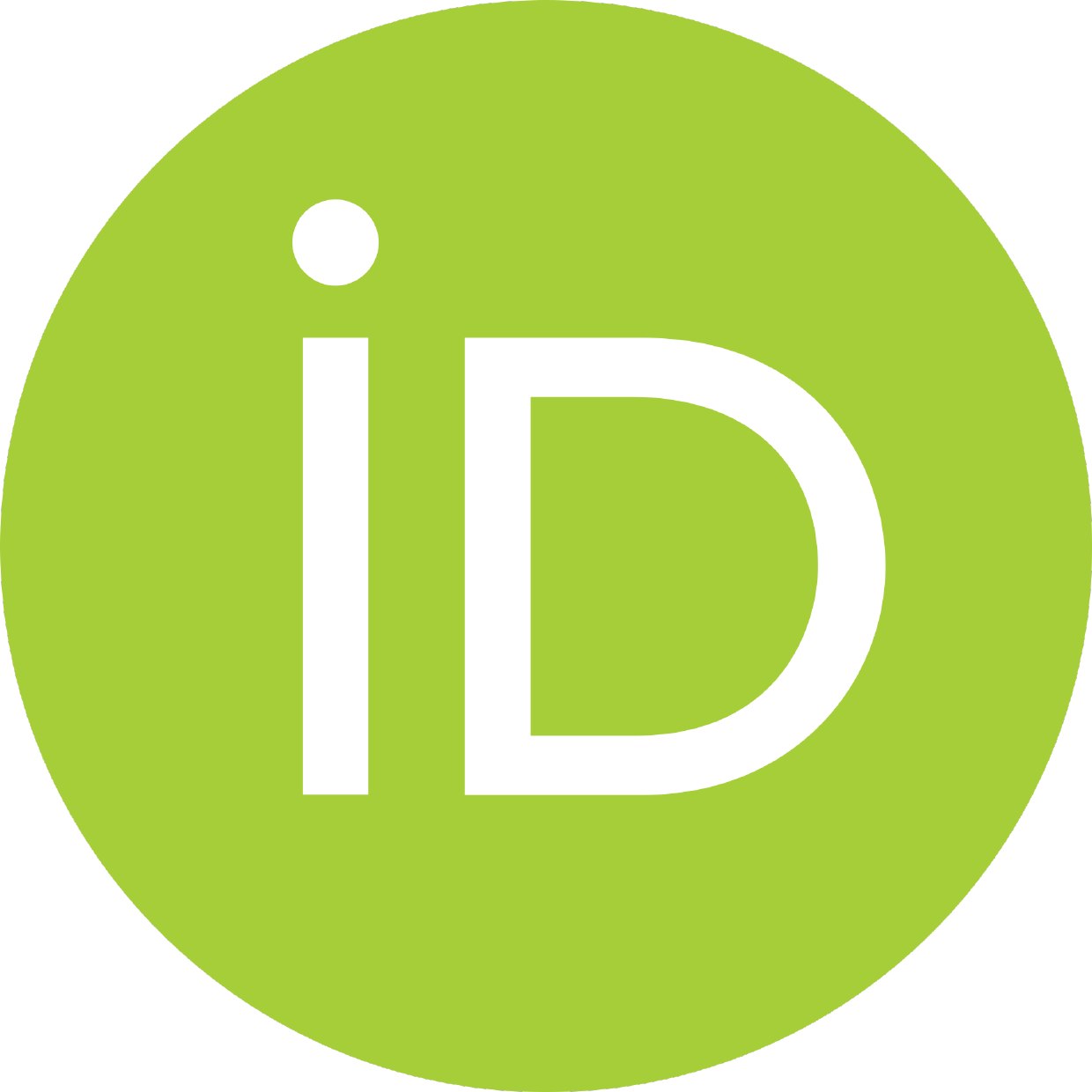}}}
\newcommand{\orcid}[1]{\href{https://orcid.org/#1}{\protect\orcidicon{#1}}}

\definecolor{seagreen}{rgb}{0.190, 0.525, 0.361}

\newcommand{\msun}{{\rm M}_\odot}
\newcommand{\petar}{\textsc{PeTar}\xspace}
\newcommand{\bseemp}{\textsc{bseEmp}\xspace}
\newcommand{\galpy}{\textsc{galpy}\xspace}

\begin{document} 

\title{\emph{Pebbles to Gems:}}
\subtitle{Intermediate-mass black holes in the first star clusters}
\authorrunning{B. Mestichelli et al.}
   \author{Benedetta Mestichelli
          \inst{1,2,3}
    \orcid{0009-0002-1705-4729} \thanks{\href{mailto:benedetta.mestichelli@gssi.it}{benedetta.mestichelli@gssi.it}}
          \and
          Manuel Arca Sedda\inst{1,2,4}\orcid{0000-0002-3987-0519}
          \and 
          Marta Volonteri \inst{5}\orcid{0000-0002-3216-1322}
          \and
          Michela Mapelli \inst{3,6,7,8} \orcid{0000-0001-8799-2548} 
          \and \\
          Stefano Torniamenti \inst{9} 
          \orcid{0000-0002-9499-1022} 
          \and
          Alessandro Lupi \inst{10,11,12}
          \orcid{0000-0001-6106-7821}
          \and 
          Marica Branchesi\inst{1,4}
          \and
          Shingo Hirano\inst{13,14}\orcid{0000-0002-4317-767X}
          \and
          Tomoaki Ishiyama \inst{15} \orcid{0000-0002-5316-9171}
          \and
          Ralf S.\ Klessen \inst{3,6}\orcid{0000-0002-0560-3172}
          \and
          Veronika Lipatova \inst{3} \orcid{0000-0002-6111-2570}
          \and
          Boyuan Liu\inst{3}\orcid{0000-0002-4966-7450}}

        \titlerunning{Intermediate-mass black holes in the first star clusters}
        \institute{
          Gran Sasso Science Institute (GSSI), Viale Francesco Crispi 7, 67100, L’Aquila, Italy
          \and
          INFN, Laboratori Nazionali del Gran Sasso, I-67100 Assergi, Italy
          \and
          Institut f\"ur Theoretische Astrophysik, Zentrum f\"ur Astronomie, Universit\"at Heidelberg, Albert Ueberle Str. 2, D-69120 Heidelberg, Germany
          \and
          INAF Osservatorio Astronomico d'Abruzzo, Via Maggini, 64100 Teramo, Italy 
          \and
          Institut d’Astrophysique de Paris, UMR 7095, CNRS and Sorbonne Universit\'e, 98 bis Boulevard Arago, 75014 Paris, France
          \and
          Universit\"{a}t Heidelberg, Interdisziplin\"{a}res Zentrum f\"{u}r Wissenschaftliches Rechnen, Im Neuenheimer Feld 225, 69120 Heidelberg, Germany
          \and
          Dipartimento di Fisica e Astronomia Galileo Galilei, Università di Padova, Vicolo dell’Osservatorio 3, I–35122 Padova, Italy
          \and
          INFN-Padova, Via Marzolo 8, I–35131 Padova, Italy
          \and Max-Planck-Institut f{\"u}r Astronomie, K{\"o}nigstuhl 17, 69117, Heidelberg, Germany   
          \and
          Como Lake Center for Astrophysics, DiSAT, Universit\`a degli Studi dell'Insubria,  via Valleggio 11, 22100, Como, Italy
          \and
          INFN, Sezione di Milano-Bicocca, Piazza della Scienza 3, I-20126 Milano, Italy
          \and
          INAF, Osservatorio Astronomico di Bologna, Via Gobetti 93/3, I-40129 Bologna, Italy    
          \and
          Kanagawa University, 3-27-1 Rokkakaku, Kanagawa-ku, Yokohama, 221-0802 Kanagawa, Japan
          \and
          The University of Tokyo, 7-3-1 Hongo, Bunkyo-ku, 113-0033 Tokyo, Japan
          \and
          Digital Transformation Enhancement Council, Chiba University, 1-33, Yayoi-cho, Inage-ku, Chiba, 263-8522, Japan
          }
    
   \date{}

   \date{}

  \abstract{
    The rapid assembly of supermassive black holes (SMBHs) observed at $z\gtrsim7$ requires efficient seeding mechanisms operating in the early Universe. Population~III (Pop.~III) star clusters have recently emerged as a promising pathway that may bridge the gap between traditional light- and heavy-seed scenarios, producing intermediate-mass black holes (IMBHs) with masses up to $\sim10^4\,\msun$. In this work, we investigate the properties and number densities of IMBHs forming in Pop.~III star clusters with masses $M_{\rm cl}\sim10^3-4\times10^5\,\msun$, and hosted in isolated dark matter minihalos, using a suite of direct $N$-body simulations. We adopt initial conditions motivated by semi-analytical cosmological models and explore a broad parameter space spanning different stellar evolution prescriptions, binary orbital parameter distributions, and cluster dynamical configurations. We find that, by $z\sim19$, the IMBH mass function consistently peaks at $m_{\rm IMBH}\sim200\,\msun$, corresponding to number densities of $n_{\rm IMBH}\sim0.2-5\,\rm cMpc^{-3}$. If Pop.~III stars formed in sufficiently dense and massive clusters, IMBHs with masses $>10^3\,\msun$ could also be produced by $z\sim19$, with number densities of $n_{\rm IMBH}\sim10^{-4}-10^{-2}\,\rm cMpc^{-3}$. The most massive IMBHs in our models reach $\sim6200\,\msun$, and originate from the collapse of very massive stars assembled through repeated stellar collisions, a process that is particularly efficient in clusters with fractal initial conditions. Lower-mass IMBHs are instead predominantly produced through single and binary stellar evolution, as well as binary stellar mergers. We further find that models combining large stellar radii and tight binaries yield the highest IMBH abundances relative to isolated Pop.~III star evolution. Owing to the high retention fraction of IMBHs ($\gtrsim88\%$), massive dense Pop.~III star clusters can act as efficient incubators of both light and heavy SMBH seeds, even if only a fraction of the overall Pop.~III stellar population formed in such environments.

  } 

   \keywords{}

   \maketitle
%

\section{Introduction}

Supermassive black holes (SMBHs) with masses $\gtrsim10^6\,\msun$ are now routinely observed at $z\gtrsim7$ \citep[e.g.,][]{greene2020, maiolino2024, akins2025, greene2026, umeda2026}. The discovery of luminous SMBHs already in place at these redshifts implies that some of them have assembled rapidly within the first $\sim1\,{\rm Gyr}$ after the Big Bang \citep[e.g.,][]{alexander2014, inayoshi2020, volonteri2021, spinoso2023, shi2024}. Explaining how such massive objects assembled on such short timescales remains one of the main challenges in models of black hole (BH) formation and galaxy evolution.

A key open question concerns the nature of the initial ``seeds'' from which SMBHs grow. Proposed formation channels are commonly divided into ``light'' and ``heavy'' SMBH seeds.

Heavy seeds are expected to form with masses $>10^3\,\msun$ through the direct collapse of massive gas clouds in atomic-cooling halos \citep[e.g.,][]{bromm2003, begelman2006, volonteri2010, regan2020, regan2024}. This scenario can naturally explain the rapid assembly of high-redshift quasars, because the seeds are already sufficiently massive to alleviate the need for prolonged super-Eddington growth. However, it requires rather specific environmental conditions to suppress molecular cooling and provide large gas inflow rates, including strong Lyman-Werner radiation backgrounds, large streaming velocities between baryons and dark matter, and halo mergers \citep[e.g.,][]{inayoshi2015, argawal2016, habouzit2016, trinca2022}.

Light seeds, instead, are typically associated with the remnants of Population~III (Pop.~III) stars, the first generation of metal-free stars formed at $z\gtrsim20$ \citep[e.g.,][]{madau2001, tanaka2009, natarajan2021, klessen2023}, or of early Population~II stars. Such seeds are expected to have masses ${\sim10^2\,\msun}$ and to be relatively abundant \citep[e.g.,][]{regan2020, tanikawa2020, tanikawa2021b, tanikawa2021, tanikawa2022,  tanikawa2023, tanikawa2024,regan2024}. However, if Pop.~III stars evolve in isolation, i.e., as single stars or in multiples that do not experience significant dynamical interactions with other stars, their remnants face several challenges as progenitors of high-redshift SMBHs. First, their relatively small masses require sustained gas accretion at, or above, the Eddington limit over extended periods \citep[e.g.,][]{volonteri2005, alexander2014, natarajan2021, shi2024}. Second, radiative and mechanical feedback from both the progenitor stars and the accreting BHs can strongly suppress gas inflow \citep[e.g.,][]{whalen2004, alvarez2009, milosavljevic2009, hosokawa2011, hirano2014, fukushima2020, liu2024b}. Finally, low-mass Pop.~III remnants may experience inefficient dynamical friction, preventing them from efficiently sinking to the centers of potential wells and thus hindering their growth via gas accretion and binary black hole (BBH) mergers \citep[e.g.,][]{pfister2019, ma2021}.

In recent years, increasing attention has been devoted to scenarios in which Pop.~III stars form in clusters rather than in isolation. Hydrodynamical simulations have shown that gas fragmentation within primordial dark matter (DM) minihalos can lead to the formation of multiple stars and compact stellar systems \citep[e.g.,][]{clark2011, greif2011, hirano2018, sugimura2020, chon2021, hirano2023}. In particular, non-zero baryon-DM streaming velocities can amplify gas asymmetries and fragmentation, potentially favoring the formation of relatively massive Pop.~III clusters \citep{hirano2018, schauer2021, hirano2023}. Dense stellar clusters provide a natural environment for the formation of intermediate-mass black holes (IMBHs), with masses $10^2-10^4\,\msun$, through repeated stellar collisions, compact object mergers, and BH-star interactions. In metal-poor or metal-free environments, where stellar winds are weak, repeated stellar collisions can efficiently produce very massive stars (VMSs), which may subsequently collapse into IMBHs \citep[e.g.,][]{pzwart2002, mapelli2016, kremer2020, rastello2020, banerjee2021, rizzuto2021, mas2023, rastello2025, paiella2026, mestichelli2026}.  

Recent studies have suggested that Pop.~III star clusters may represent an efficient channel for the formation of SMBH seeds with masses $\gtrsim10^3\,\msun$. Several works have explored the impact of runaway collisions, BBH mergers, and hierarchical cluster assembly on the growth of primordial BHs \citep[e.g.,][]{wang2022, mestichelli2024, reinoso2025, vergara2025, rantala2026b, wu2026}. These studies indicate that stellar dynamics may bridge the gap between traditional light- and heavy-seed scenarios, producing IMBHs with masses $\sim10^3-10^4\,\msun$ already at very high redshifts. Nevertheless, the initial conditions of Pop.~III star clusters remain highly uncertain, including their mass and density, as well as the initial mass function (IMFs) and the orbital parameter distributions of primordial binaries. 

Here, we address these uncertainties by performing a suite of direct $N$-body simulations based on cosmologically motivated initial conditions and exploring a broad range of stellar evolution models and cluster configurations. We extract cluster masses and DM minihalo properties from the semi-analytical cosmological models of \citet{hartwig2022} and \citet{ishiyama2025}, and evolve the clusters with the hybrid $N$-body code \petar \citep{wang2020b}, including prescriptions for single and binary stellar evolution through \bseemp \citep{tanikawa2020}, and assuming high primordial binary fractions \citep{moe2017}. We explore the impact of different stellar evolution tracks, binary orbital parameters, and initial dynamical configurations. Our main goal is to assess whether stellar dynamics in the first star clusters can efficiently produce IMBHs that may act as SMBH seeds. In particular, we study the relative importance of stellar evolution, stellar collisions, VMS formation, and BBH mergers in shaping the mass spectrum and number density of dynamical IMBHs.

The paper is structured as follows. Section~\ref{sec:methods} presents the initial conditions and outlines the used numerical code. In Sec.~\ref{sec:results}, we describe our main results. Section~\ref{sec:discussions} considers the impact of our chosen prescriptions and discusses the main implications of our results on SMBH seeding. Finally, Sec.~\ref{sec:summary} summarizes our conclusions.

\section{Methods}\label{sec:methods}

In this work, we simulate Pop.~III star clusters forming at the center of DM minihalos, whose properties are extracted from the semi-analytical cosmological simulations of \citet[][henceforth \citetalias{hartwig2022}]{hartwig2022} and \citet[][\citetalias{ishiyama2025} from here forward]{ishiyama2025}.

Our models consist of two sets of $N$-body simulations based on these cosmologically motivated initial conditions. We performed the simulations using the $N$-body code \petar \citep{wang2020b}, which includes prescriptions for single and binary stellar evolution through population-synthesis algorithms \citep[\bseemp;][]{tanikawa2020}, as well as for external tidal effects \citep[\galpy;][]{bovy2015}.

\subsection{Cosmological simulations}\label{sec:methods_cosmo}

To initialize our cluster models, we extracted from \citetalias{hartwig2022} and \citetalias{ishiyama2025} the masses of both the star clusters and the host minihalos at $z\sim20$. We assumed that the initial cluster mass, $M_{\rm cl}$, corresponds to the total Pop.~III stellar mass formed within each minihalo. 

\begin{figure*}[ht]
    \centering
    \includegraphics[width=0.95\textwidth]{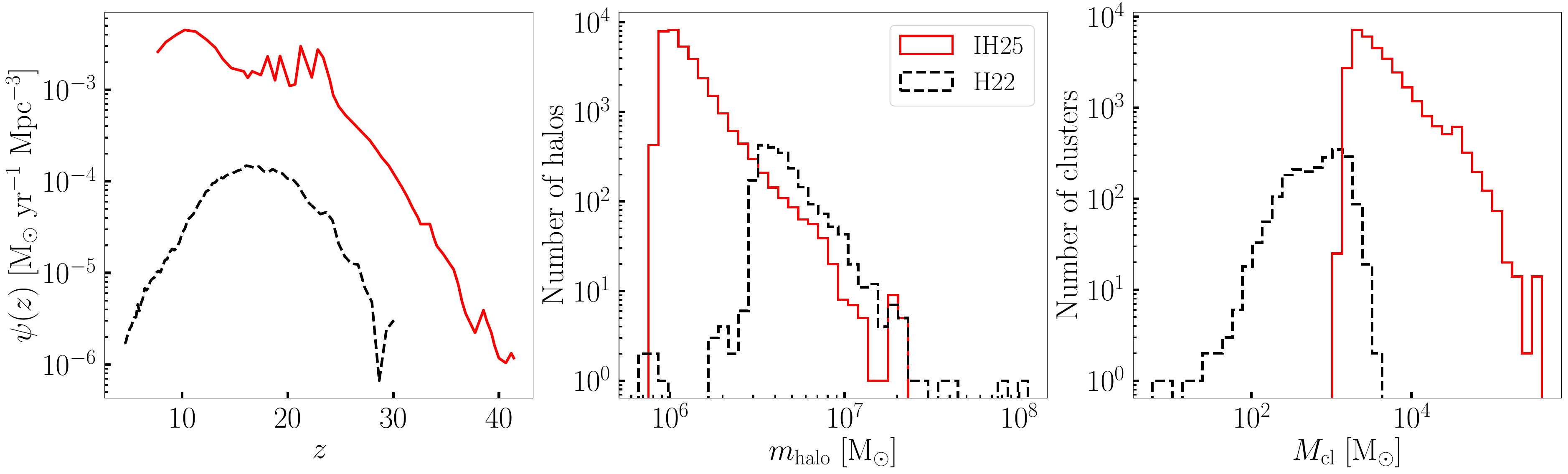}
    \caption{Star formation rate density, halo mass distribution, and cluster mass distribution obtained from \citetalias{hartwig2022} (black dashed line) and \citetalias{ishiyama2025} (red continuous line). The halo and cluster mass distributions are shown at $z\sim20$.}
     \label{fig:sf_halo_cl}
\end{figure*}

Figure~\ref{fig:sf_halo_cl} shows the star formation rate densities (SFRDs) and the distribution of masses of star-forming DM minihalos and clusters at $z\sim20$, as derived from \citetalias{hartwig2022} and \citetalias{ishiyama2025}. We started our simulations at this redshift because the Pop.~III SFRD is close to its peak while still dominating over the Population~II contribution, and the DM minihalos are massive enough to host a star cluster. We notice that, although the star-forming halos in \citetalias{ishiyama2025} are systematically less massive than those in \citetalias{hartwig2022}, their hosted clusters are typically more massive in the former case. The origin of this difference is discussed below.

\subsubsection{H22}

\citet{hartwig2022} introduce the semi-analytical framework implemented in the code \textsc{A-sloth} \citep[see also][]{magg2022, hartwig2024}. The model is based on DM halo merger trees constructed within a comoving volume of $L=8\,h^{-1}\,\rm Mpc$, onto which analytical prescriptions for the formation of individual metal-free and metal-poor stars are applied \citep{ishiyama2016, griffen2016, griffen2018}. The critical halo mass for star formation is defined as the minimum between $M_{\rm crit}$ from \citet{schauer2021} and the mass threshold for atomic hydrogen cooling \citep{hummel2012}. Only halos with virial masses above this threshold can form Pop.~III stars. The \textsc{A-sloth} framework assumes a constant star formation efficiency, with a best-fit value $\eta_{\rm SFE}=0.38$. It samples single stars with a log-flat \citep[$\xi(m)\propto m^{-1}$, e.g.,][]{sb2013} IMF and includes radiative, chemical, and mechanical feedback processes. A global Lyman-Werner background is included, although it is not treated self-consistently. The baryon-DM streaming velocity adopted in \citetalias{hartwig2022} is $v_{\rm BC}=0.8\,\sigma_{\rm BC}$, with $\sigma_{\rm BC}=30\,\rm km\,s^{-1}$ at recombination. The model is calibrated against a wide range of observables from both the local and high-redshift Universe, and it adopts cosmological parameters from \citet{planck2016}.

\subsubsection{IH25}

\citet{ishiyama2025} describe a semi-analytical framework that also operates on DM halo merger trees, generated within a comoving volume of $L=16\,h^{-1}\,\rm Mpc$ \citep{ishiyama2021}. In this model, the critical halo mass is defined as the minimum between $M_{\rm crit}$ from \citet{kulkarni2021} and the threshold for atomic cooling from \citet{feathers2024}. Pop.~III star formation depends explicitly on halo properties, such as the gas infall rate at the Jeans scale, following the results of hydrodynamical simulations \citep{hirano2014, hirano2015}. Moreover, the model takes into account the enhanced star formation efficiency in rapidly collapsing clouds \citep[$>0.01\,\msun\rm\,yr^{-1}$,][]{toyouchi2023}. The Lyman-Werner radiation field is modeled self-consistently, leading to a time-dependent regulation of star formation \citep{johnson2013}. Once the halo mass exceeds the critical mass for star formation, it is assumed that a single star forms instantaneously in the halo. It is important to note that non-zero streaming velocities can enhance gas fragmentation, increasing the likelihood of forming multiple Pop.~III stars in a minihalo \citep{hirano2018, hirano2023}. Therefore, in this work we adopt the results obtained in \citetalias{ishiyama2025} with a baryon-DM streaming velocity of $v_{\rm BC}=1\,\sigma_{\rm BC}$, to remain consistent with \citetalias{hartwig2022}. The model assumes cosmological parameters from \citet{aghanim2020}.

As shown in Fig.~\ref{fig:sf_halo_cl}, the different treatment of star formation efficiency and radiative feedback results in systematically larger Pop.~III stellar masses and, consequently, more massive clusters, than in \citetalias{hartwig2022}, for similar halo masses. In addition, the different prescription for the critical halo mass allows Pop.~III star formation to occur in smaller halos. 

\subsection{$N$-body code}
We performed the simulations using the hybrid $N$-body code \petar \citep{wang2020b}. Hybrid schemes rely on Hamiltonian splitting of long-range and short-range interactions. In particular, \petar implements the particle-tree particle-particle \citep[P$^3$T;][]{oshino2011} algorithm together with the slow-down algorithmic regularization method \citep[SDAR;][]{wang2020a} to accurately and efficiently treat close encounters and binary interactions. The use of OpenMP for parallel workload distribution across CPUs, combined with GPU acceleration, ensures high computational efficiency and scalability. The code includes prescriptions for single and binary stellar evolution via the population-synthesis code \bseemp \citep{tanikawa2020}, while the effect of the host DM halo is modeled using the \galpy package \citep{bovy2015}.

\subsubsection{Compact object mergers in \petar}
In dense star clusters with high primordial binary fractions, BH-star collisions can occur frequently as a result of the combined effects of binary evolution and dynamical interactions \citep[e.g.,][]{mapelli2016,mas2024, rastello2025}. The outcome of such interactions remains uncertain \citep[e.g.,][]{kremer2020, schroder2020}. Nonetheless, dynamical studies \citep{kiroglu2025, kiroglu2025b} indicate that repeated BH-star collisions may drive significant BH growth. In this work, we assumed that a fraction $f_{\rm c}=0.5$ of the stellar mass involved in such interactions is accreted onto the BH \citep{banerjee2021, rizzuto2021, mas2024}.

Gravitational-wave (GW) driven mergers are modeled by evolving the semi-major axis and eccentricity of binaries following \citet{peters1964}. A binary is considered to merge when its GW inspiral timescale becomes shorter than its integration timestep. At that point, the system is evolved to the merger time, allowing us to record the position of each BBH merger within the cluster.

\subsubsection{External potential}

We modeled the external gravitational potential using the \galpy package \citep{bovy2015}, assuming that each star cluster resided at the center of its host DM minihalo.  For \citetalias{hartwig2022}, we randomly selected 18 star clusters and their corresponding host DM minihalos. For \citetalias{ishiyama2025}, instead, we selected one cluster and its host minihalo from each logarithmic cluster-mass bin, for a total of 22 systems. For each halo, we computed the virial radius using Eq.~24 of \citet{barkana2001}. We modeled the DM potential with a Navarro-Frenk-White profile \citep{nfw1996}, adopting a constant concentration parameter $c=3.5$, consistent with \citet{correa2015}. We assumed a simulation time of $20\,\rm Myr$, ensuring consistency with the assumption that the host minihalos evolve in isolation during this timescale. The simulation time is always larger than or comparable to the half-mass relaxation time of the clusters \citep{spitzer1988}. 

\subsection{Stellar and binary evolution}
We modeled single and binary stellar evolution using the population-synthesis code \bseemp \citep{tanikawa2020}. All clusters have metallicity $Z = 10^{-10}$. Binary evolution processes, including tides, mass transfer, common-envelope evolution, and GW-driven orbital decay follow the prescriptions of \citet{hurley2002}. We treated the common-envelope phase using the standard $\alpha_{\rm CE}-\lambda_{\rm CE}$ formalism, where $\alpha_{\rm CE}$ is the fraction of orbital energy used to unbind the envelope and $\lambda_{\rm CE}$ characterizes the envelope binding energy. We adopted $\alpha_{\rm CE} = 1$, while $\lambda_{\rm CE}$ was computed following \citet{claeys2014}.

\subsubsection{Stellar tracks}
The \bseemp code provides two sets of stellar evolution fitting formulae, referred to as the ``M'' and ``L'' tracks from the \textsc{hoshi} code \citep{takahashi2016,yoshida2019}, calibrated using observations of stars in the Milky Way and the Large Magellanic Cloud, respectively. The two models differ primarily in the efficiency of convective overshooting, with overshoot parameters $f_{\rm ov}=0.01$ for the M model and $f_{\rm ov}=0.03$ for the L model \citep{tanikawa2022}. For Pop.~III stars, these differences lead to a markedly distinct mass-radius relation. M-tracks stars reach radii of at most $\sim100\,\rm R_{\odot}$ even for $160\,\msun$, whereas L-tracks stars can exceed $10^3\,\rm R_{\odot}$ for zero-age main sequence (ZAMS) masses $m_{\rm ZAMS}\geq80\,\msun$. The enhanced overshooting in the L model enables more efficient mixing of fresh hydrogen into the core, producing helium cores that are $\sim20\%$ more massive at the end of the main sequence. This results in higher core luminosities and substantially larger post-main-sequence radii, and has an impact on the (pulsational) pair-instability supernova (PISN) mass range.

We sampled stars with ZAMS masses between $0.08$ and $300\,\msun$ from a log-flat IMF \citep[$\xi(m)\propto m^{-1}$,][]{sb2013, stacy2016, hirano2014, hirano2015, susa2014, wollenberg2020, chon2021, tanikawa2021, jaura2022, prole2022, klessen2023}. We point out that Pop.~III clusters of different masses and formed in different environments can have different stellar IMFs, as shown in some hydrodynamic simulations and analytical models \citep{liu2024b, prole2024}. Here, we adopt an invariant Pop.~III IMF for simplicity. In the following, we define a VMS as a star with $m_*>300\,\msun$, and an IMBH as a BH with $m\geq100\,\msun$.

\subsubsection{Compact remnants in \bseemp}
For core-collapse supernovae we adopted the rapid explosion model by \citet{fryer2012},  yielding BHs with a mass $>5\,\msun$. Natal kicks for BHs were assigned according to
\begin{equation}
v_{\rm BH} = (1 - f_{\rm fb})\, v_{\rm NS},
\end{equation}
where $v_{\rm NS}$ is the natal kick velocity of neutron stars, drawn from the distribution of \citet{phinney1992}, and $f_{\rm fb}$ is the fallback fraction. The parameter $f_{\rm fb}$ ranges from 0 to 1, with $f_{\rm fb}=1$ corresponding to direct collapse and a zero natal kick.

Pair-instability supernovae were treated following the prescriptions by \citet{belczynski2016}. Stars with final helium core masses $40 \le m_{\rm He,f} < 65\,\msun$ undergo pulsational PISNe, while full PISNe occur for $65 \le m_{\rm He,f} \le 135\,\msun$. Stars with $m_{\rm He,f} > 135\,\msun$ collapse directly into BHs. The ZAMS mass ranges corresponding to the PISN gap are therefore $120 \lesssim m_{\rm ZAMS} \lesssim 240\,\msun$ for L tracks and $135 \lesssim m_{\rm ZAMS} \lesssim 250\,\msun$ for M tracks.

For pulsational PISNe, we adopted the prescription of \citet{leung2019}, corresponding to the ``moderate'' case of \citet{belczynski2020}. The final remnant mass is given by
\begin{equation}
\begin{cases}
0.65\,m_{\rm He,f} + 12.2 & 40 \le m_{\rm He,f} < 60\,\msun, \\
51.2 & 60 \le m_{\rm He,f} < 62.5\,\msun, \\
-14.3\,m_{\rm He,f} + 938 & 62.5 \le m_{\rm He,f} \le 65\,\msun.
\end{cases}
\end{equation}

As a result, the BH mass gap in the L and M tracks corresponds to $60 \lesssim m \lesssim 170\,\msun$ and $80 \lesssim m \lesssim 200\,\msun$, respectively. These differences arise from the larger helium core masses produced at a fixed stellar mass in the L model.

\subsection{Initial conditions}
We generated the initial conditions of our simulations using \textsc{mcluster} \citep{kuepper2011}. We extracted cluster masses from \citetalias{hartwig2022} and \citetalias{ishiyama2025}, assuming that the total Pop.~III stellar mass forming in a halo at $z\sim20$ corresponds to the initial cluster mass. As shown in Fig.~\ref{fig:sf_halo_cl}, both distributions peak at $\sim10^3\,\msun$, but \citetalias{ishiyama2025} also produces a large fraction of clusters up to $\sim4\times10^5\,\msun$. We simulated clusters with $M_{\rm cl} \in [10^3,3\times10^3]\,\msun$ ($N\in[50, 80]$) for \citetalias{hartwig2022} and $M_{\rm cl} \in [10^3,4\times10^5]\,\msun$ ($N\in [50, 1.5\times10^4]$) for \citetalias{ishiyama2025}.

Positions and velocities of particles in the clusters were sampled from a King profile \citep{king1966} with central potential $W_0=5$, a commonly adopted fiducial value corresponding to a moderately concentrated cluster \citep{pzwart2004, giersz2006, kremer2020}. We assumed two values for the half-mass radius, $r_{\rm h}=0.5$ and $1\,\rm pc$, consistent with \citet{marks2012}. We note that this relation is applicable in the local Universe, but it has not been specifically established for Pop.~III star clusters. However, \citet{liu2024b} find a relation connecting the initial half-mass radius of Pop.~III clusters to their mass, which reproduces values for $r_{\rm h}$ that are consistent or smaller than the ones used in this work. 

\begin{figure*}[ht] 
\centering 
\includegraphics[width=0.9\textwidth]{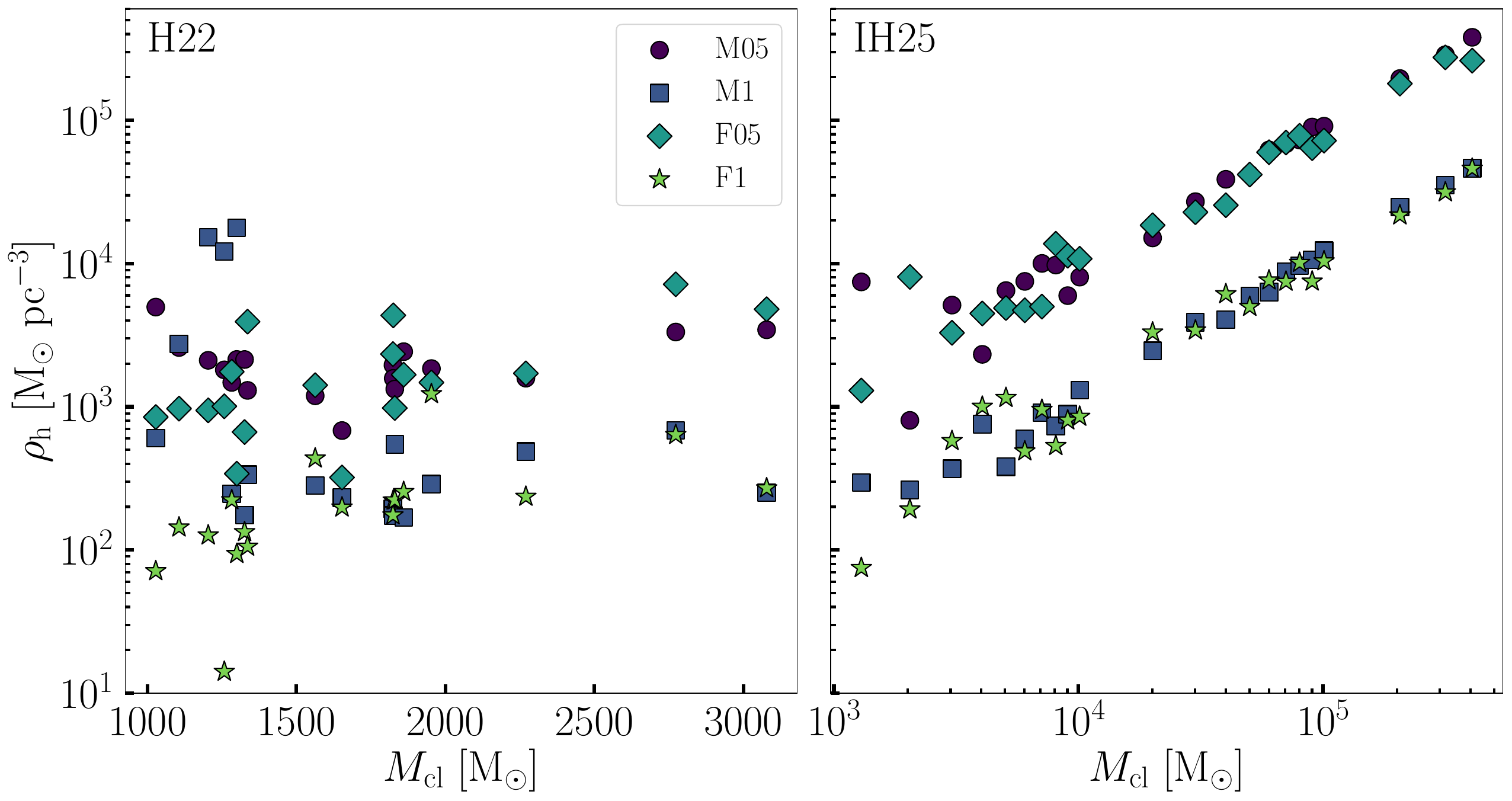} 
\caption{Initial density at half-mass radius $\rho_{\rm h}$ as a function of the initial cluster mass $M_{\rm cl}$ for the two cosmological sets (\citetalias{hartwig2022} on the left and \citetalias{ishiyama2025} on the right) and for the four dynamical configurations (different markers): monolithic with $r_{\rm h}=0.5\,\rm pc$ (M05), monolithic with $r_{\rm h}=1\,\rm pc$ (M1), fractal with $r_{\rm h}=0.5\,\rm pc$ and $D=1.6$ (F05), and fractal with $r_{\rm h}=1\,\rm pc$ and $D=1.6$ (F1).}
\label{fig:init_cond} 
\end{figure*}

We considered both monolithic and fractal initial conditions, adopting a fractal dimension $D=1.6$ to capture the expected substructure of star-forming regions \citep[e.g.,][]{sanchez2009, kuepper2011, dicarlo2019}. We denote our models as M05 (monolithic, $r_{\rm h}=0.5\,\rm pc$), M1 (monolithic, $r_{\rm h}=1\,\rm pc$), F05 (fractal, $r_{\rm h}=0.5\,\rm pc$), and F1 (fractal, $r_{\rm h}=1\,\rm pc$). Figure~\ref{fig:init_cond} shows the initial densities at the half-mass radius $\rho_{\rm h}$ as a function of cluster mass for the two cosmological models and the four dynamical configurations. We notice that, while $\rho_{\rm h}$ is the same for monolithic and fractal clusters, the local density in fractal substructures can be higher.

We modified the properties of the primordial binaries using \textsc{raccoonBinator}\footnote{\url{https://github.com/bmestichelli/RaccoonBinator.git}} to be in accordance with observations in the local Universe and simulations of Pop.~III stars, expanding the model presented by \citet{torniamenti2021}. The primordial binary fraction depends on the primary mass following \citet{moe2017}. The global binary fraction is $f_{\rm b}=N_{\rm b}/(N_{\rm s}+N_{\rm b})\simeq0.23$, corresponding to $\sim38\%$ of stars in binaries. We drew binary mass ratios, orbital periods, and eccentricities from the distributions proposed by 
\citet[][hereafter \citetalias{sana2012}]{sana2012}, based on observations of nearby O- and B-type stars, and by \citet[][hereafter \citetalias{sb2013}]{sb2013}, based on cosmological simulations of Pop.~III stars. The mass-ratio distributions ($q=m_2/m_1$) are given by
\begin{equation}
\begin{cases}
\xi(q)\propto q^{-0.1} & \rm S12, \\
\xi(q)\propto q^{-0.55} & \rm SB13,
\end{cases}
\end{equation}
with $q\in[0.01,1]$. Compared to \citetalias{sana2012}, the \citetalias{sb2013} prescription therefore favors binaries with smaller mass ratios. The orbital period distributions follow
\begin{equation}
\begin{cases}
\xi(\Pi)\propto \Pi^{-0.55} & \rm S12, \\
\xi(\Pi)\propto \exp\!\left[-(\Pi-\mu)^2/(2\sigma^2)\right] & \rm SB13,
\end{cases}
\end{equation}
where $\Pi=\log(P/{\rm day})$. For \citetalias{sana2012}, we adopted $\Pi\in[0.15,5.5]$, extending the maximum orbital period to $\sim10^5$ days in order to include wide binaries. For \citetalias{sb2013}, we instead assumed $\mu=5.5$ and $\sigma=0.85$. As a consequence, the \citetalias{sb2013} distribution typically produces wider binaries than \citetalias{sana2012}. The eccentricity distributions are
\begin{equation}
\begin{cases}
\xi(e)\propto e^{-0.42} & \rm S12, \\
\xi(e)\propto 2e & \rm SB13,
\end{cases}
\end{equation}
with $e\in[0,1)$.  \citetalias{sana2012}  therefore favor lower eccentricities compared to the thermal distribution adopted by \citetalias{sb2013}.

Each simulation based on \citetalias{hartwig2022} initial conditions was repeated ten times, while those based on \citetalias{ishiyama2025} were repeated five times, owing to their higher computational cost. A summary of the adopted initial conditions is provided in Table~\ref{table:ic_sims}.

\begin{table*}[ht!] 
\centering 
\caption{Initial conditions of the sets of simulations.} 
\begin{tabular}{c c c c c c c} 
\hline 
Cosmological IC & $M_{\rm cl}$ & $r_{\rm h}$ & $\rho_{\rm h}$ & Repetitions & Dynamical configuration & Orbital parameters\\ [0.5ex]
& $\msun$ & $\rm pc$ & $\msun\,\rm pc^{-3}$ & & &\\ [0.5ex] 
\hline\hline 
\citetalias{hartwig2022} & $1000-3000$ & 0.5,1 & $10-10^4$ & 10 & Monolithic, Fractal ($D=1.6$) & \citetalias{sana2012}, \citetalias{sb2013}\\ 
\citetalias{ishiyama2025} & $1000-4\times10^5$ & 0.5,1 & $80-5\times10^5$ & 5 & Monolithic, Fractal ($D=1.6$) & \citetalias{sana2012}, \citetalias{sb2013}\\ 
\hline \end{tabular} 
\tablefoot{Column 1: set of cosmological initial conditions (ICs). Column 2: initial cluster mass range. Column 3: initial half-mass radii. Column 4: range of the initial density at half-mass radius. Column 5: number of repetitions of each model. Column 6: initial dynamical configurations. Column 7: distributions of orbital parameters.} 
\label{table:ic_sims} 
\end{table*}

\section{Results}\label{sec:results}

\subsection{Number density of IMBHs}

\begin{figure*}[ht]
    \centering
    \includegraphics[width=\textwidth]{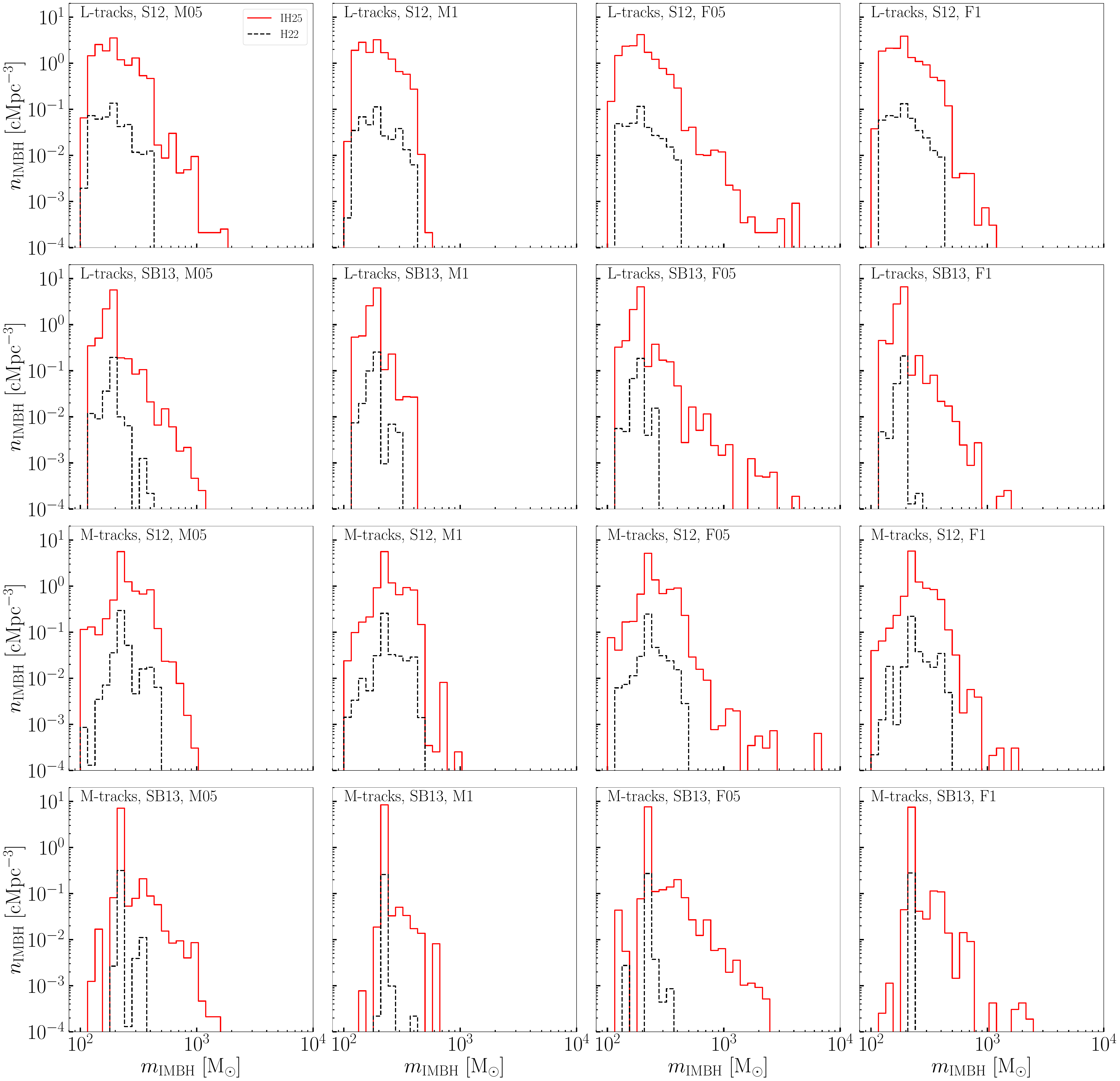}
    \caption{Number density of IMBHs per logarithmic mass bin at $z\sim19$, weighted by the cluster mass distribution at $z\sim20$. The red continuous line shows the number density for \citetalias{ishiyama2025}, while the black dashed line shows the same distribution for \citetalias{hartwig2022}. From left to right, different star cluster models: M05 (monolithic, $r_{\rm h}=0.5\,\rm pc$), M1 (monolithic, $r_{\rm h}=1\,\rm pc$), F05 (fractal, $r_{\rm h}=0.5\,\rm pc$), and F1 (fractal, $r_{\rm h}=1\,\rm pc$). The two upper rows adopt the L tracks (large overshooting), whereas the two lower rows adopt the M tracks (small overshooting). S12 and SB13 indicate initial orbital parameters of binary stars from \cite{sana2012} and \cite{sb2013}, respectively.}
     \label{fig:n_seed}
\end{figure*}

Figure~\ref{fig:n_seed} shows the number density distributions of IMBHs at $z\sim19$, originating from star clusters that formed at $z\sim20$ and subsequently evolved for $20\,\rm Myr$. We present all model configurations and consider the two sets of cosmological initial conditions (\citetalias{hartwig2022, ishiyama2025}). We compute the cosmologically-weighted number density of IMBHs by weighting each of them by the cosmological abundance of its host cluster. For each IMBH formed in a cluster of mass $M_{\rm cl}$, we assign a cosmological weight $w = n_{\rm{III}}(M_{\rm{cl}}) \, \Delta\log_{10}M_{\rm cl} / N_{\rm{sim}}(M_{\rm{cl}})$, where $n_{\rm III}(M_{\rm cl})$ is the comoving number density of clusters with mass $M_{\rm cl}$ per dex of $M_{\rm cl}$ from the cosmological simulations \citepalias{hartwig2022, ishiyama2025}, $\Delta\log_{10}M_{\rm cl}$ is the logarithmic width of the cluster mass bins, and $N_{\rm sim}(M_{\rm cl})$ is the number of times that cluster mass was sampled across replications. This weighting scheme accounts for the cosmic abundance of each cluster mass, yielding the expected comoving number density of IMBHs as a function of their mass, $n_{\rm IMBH}(m_{\rm IMBH})$ [cMpc$^{-3}$]. We assume that only first-generation BBH mergers could produce IMBHs (see Sec.~\ref{sec:caveats}). Moreover, we include IMBHs that are expelled from their host cluster through dynamical interactions by the end of the simulation, even though they are unlikely to represent SMBH seeds.

For $m_{\rm IMBH}<200\,\msun$, different assumptions on stellar evolution and orbital parameters result in noticeably different IMBH mass distributions. In fact, the PISN mass range is broader on the M tracks than on L tracks. Moreover, L-tracks stars expand more and lose more mass, leaving lighter remnants. Binary interactions can narrow this effective range, especially with the smaller initial semi-major axes from \citetalias{sana2012}, which boost mass transfer and mergers. 

In all configurations, the IMBH mass distribution peaks at $m_{\rm IMBH}\sim200\,\msun$, corresponding to number densities of $n_{\rm IMBH}\sim0.2\,\rm cMpc^{-3}$ for \citetalias{hartwig2022} and $n_{\rm IMBH}\sim5\,\rm cMpc^{-3}$ for \citetalias{ishiyama2025}. This peak arises because the dominant formation channel for IMBHs in our simulations is the direct collapse of stars with masses above the PISN mass gap. Such stars can form through single and binary stellar evolution, binary mergers or, less frequently, through short chains of stellar collisions. The peak at $m_{\rm IMBH}\sim200\,\msun$ is produced both in \citetalias{hartwig2022} and in \citetalias{ishiyama2025}, despite the different cluster mass distributions (Fig.~\ref{fig:sf_halo_cl}). The two models, in fact, favor the formation of small clusters, while larger IMBHs can be produced only in dense clusters with short relaxation times (see Sec.~\ref{sec:formation_channels}). We will discuss the impact of the chosen IMF and maximum mass on this result in Appendix~\ref{app:imf_impact}.

For $m_{\rm IMBH}\lesssim 500\,\msun$, models based on \citetalias{hartwig2022} predict number densities over an order of magnitude lower than those using \citetalias{ishiyama2025}. This mainly reflects (i) the larger cluster masses in \citetalias{ishiyama2025}, which form more IMBHs, and (ii) their lower critical mass for Pop.~III star formation, allowing stars to form in smaller halos.

Different cluster mass distributions produce large discrepancies at the high-mass end. The larger cluster masses in \citetalias{ishiyama2025} yield IMBHs up to an order of magnitude more massive than those from \citetalias{hartwig2022}, which reach $m_{\rm IMBH}\sim500\,\msun$ for \citetalias{sana2012} and $\lesssim400\,\msun$ for \citetalias{sb2013}. With initial conditions from \citetalias{ishiyama2025}, the most massive IMBHs form in initially fractal configurations (F05, F1), where higher densities are reached during the first $\sim2\,\rm Myr$ than in the monolithic case (see Sec.~\ref{sec:formation_channels}). Here, IMBHs can exceed $2000\,\msun$, with masses up to $\sim6200\,\msun$ (M-tracks, \citetalias{sana2012}, F05), and number densities $n_{\rm IMBH}\sim{\rm few}\times10^{-4}-10^{-3}\,\rm cMpc^{-3}$. Because of their lower density, none of the M1 models produce IMBHs with $m_{\rm IMBH}>10^3\,\msun$. 

Overall, our results suggest that, if Pop.~III stars commonly formed in clusters with $M_{\rm cl}\gtrsim10^3\,\msun$, they could produce IMBHs with masses up to $300-500\,\msun$ at number densities comparable to, or exceeding, those expected for light seeds from isolated Pop.~III stars \citep[e.g.,][]{madau2001, regan2020, regan2024, mehta2026, prole2026}. Moreover, if dense massive Pop.~III clusters were to form, they would also create a non-negligible population of IMBHs with masses above $10^3\,\msun$ already at $z\sim19$. In Sec~\ref{sec:cfr_lit} we will compare our current results with previous studies on SMBH seeds, while in Sec.~\ref{sec:seeds_growth} we will discuss the potential to grow of our IMBHs.

\subsection{Formation and growth channels of IMBHs}\label{sec:formation_channels}

Table~\ref{table:form_chan} reports the median contribution of the different IMBH formation channels. We define the channels as: i) stellar evolution (SE), if the IMBH derives from the collapse of a star with mass $\leq300\,\msun$, which formed via single and binary stellar evolution, stellar mergers or collisions; ii) VMS, if the IMBH derives from the collapse of a star with $m_*>300\,\msun$; iii) BBHm, if the IMBH forms via a single BBH merger. Across all masses and configurations, the dominant formation channel ($\gtrsim65\%$) is SE. This channel is particularly important ($\gtrsim95\%$) in simulations with wide binaries \citepalias{sb2013}, where IMBHs predominantly originate from the direct collapse of single stars above the PISN gap. In simulations adopting orbital parameters from \citetalias{sana2012}, instead, up to $\sim35\%$ of IMBHs form through the collapse of VMSs. Only in the most massive clusters (\citetalias{ishiyama2025} with $M_{\rm cl}>3000\,\msun$), a small fraction of IMBHs (up to $0.1\%$) originates from BBH mergers. 

The properties of the most massive IMBHs formed in our simulations can be found in Appendices~\ref{app:h22_maxseed} and \ref{app:maximum_seeds}. Above $M_{\rm cl}\gtrsim40,000\,\msun$, these IMBHs are preferentially produced in clusters with initial fractal conditions, both for \citetalias{sana2012} and \citetalias{sb2013} (see Appendix~\ref{app:maximum_seeds}). This is a consequence of the higher densities reached in these systems during the first million years of cluster evolution. Given the chosen log-flat IMF and upper stellar mass limit ($m_{\rm max}=300\,\msun$), stellar collision chains can yield VMSs with masses up to $\sim7000\,\msun$ that subsequently form IMBHs via direct collapse or a collision with a BH \citep[e.g.,][]{pzwart2002, rizzuto2021, mas2023, rantala2024, rantala2025,rantala2026, paiella2026, mestichelli2026}. We find that the properties of the repeated stellar collisions depend on the adopted binary population. 

In models with \citetalias{sana2012}, binaries are predominantly hard and are not easily ionized. As a result, the first collisions occur early, often involving primordial binaries located in dense substructures, where the local stellar density is high. 

\begin{figure}[ht]
    \centering
    \includegraphics[width=0.9\columnwidth]{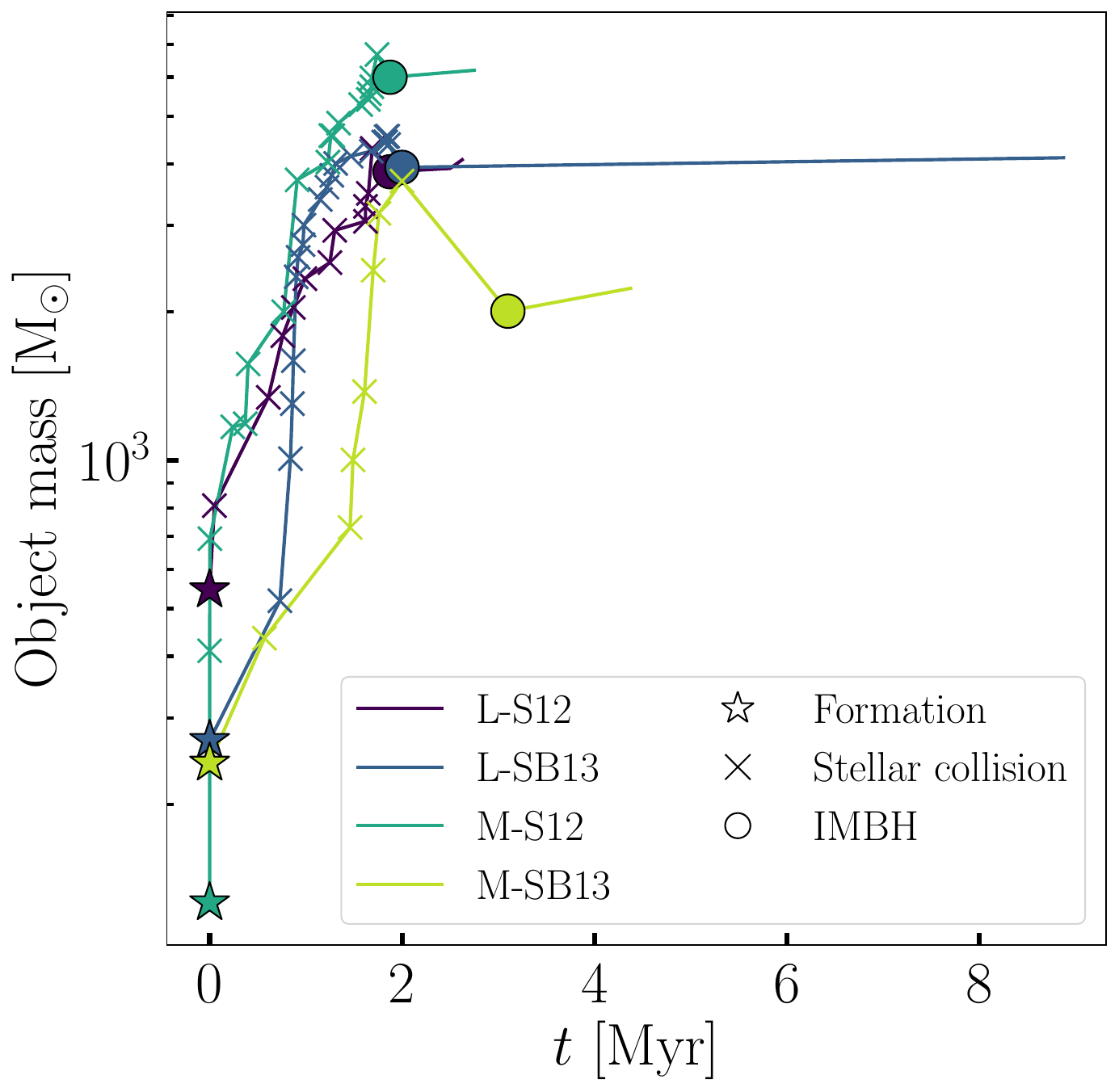}
    \caption{Example of mass growth tracks leading to the formation of the most massive IMBHs in the four configurations, assuming \citetalias{ishiyama2025} and F05.}
     \label{fig:imbh_growth_track}
\end{figure}

In contrast, in models adopting \citetalias{sb2013}, binaries are typically soft and are quickly disrupted. Consequently, most collisions are dynamical in origin and occur at later times, as shown in  Fig.~\ref{fig:imbh_growth_track}. In this case, the onset of collisions is further facilitated by the reduced expansion of the cluster, as the lack of hard binaries limits dynamical heating.

The tracks shown in Fig.~\ref{fig:imbh_growth_track} all correspond to IMBHs forming from the direct collapse of a VMS. The final IMBHs have a larger mass because of accretion from a companion star or a BBH merger. In general, we find that $\lesssim10\%$ ($\lesssim1\%$) IMBHs forming via SE or VMS collapse can grow further thanks to stellar collisions (a BBH merger). Out of the ones undergoing stellar collisions, only $\lesssim3\%$ will accrete $>5\,\msun$. Finally, $\lesssim12\%$ IMBHs (with masses usually below $\sim 500\,\msun$) escape their parent clusters because of BBH mergers or dynamical interactions. 

\setlength{\tabcolsep}{5pt}
\renewcommand{\arraystretch}{1.1}

\begin{table}[ht!]
\centering
\caption{IMBHs formation channels.}
\begin{tabular}{c c c c c} 
 \hline
 Cosmology & Model & SE & VMS & BBHm\\  
 & & $\%$ & $\%$ & $\%$\\ 
 \hline\hline 
 \multirow{4}{*}[-2pt]{\citetalias{hartwig2022}}
                   & L-tracks, \citetalias{sana2012} & 79.1 & 20.9 & 0\\
                   & M-tracks, \citetalias{sana2012} & 84.3 & 15.7 & 0\\
                   & L-tracks, \citetalias{sb2013} & 98.1 & 1.9 & 0\\
                   & M-tracks, \citetalias{sb2013} & 100 & 0 & 0\\
 \hline\hline 
 \multirow{4}{*}[-2pt]{\shortstack{\citetalias{ishiyama2025} \\ $M_{\rm cl}\leq3000\,\msun$}}
                   & L-tracks, \citetalias{sana2012} & 65  & 35 & 0\\
                   & M-tracks, \citetalias{sana2012} & 73 & 27 & 0\\
                   & L-tracks, \citetalias{sb2013} & 100 & 0 & 0\\
                   & M-tracks, \citetalias{sb2013} & 100 & 0 & 0\\
 \hline\hline 
 \multirow{4}{*}[-2pt]{\shortstack{\citetalias{ishiyama2025} \\ $M_{\rm cl}>3000\,\msun$}}
                   & L-tracks, \citetalias{sana2012} & 72.9 & 27.1 & 0\\
                   & M-tracks, \citetalias{sana2012} & 70 & 29.9 & 0.1\\
                   & L-tracks, \citetalias{sb2013} & 95.5 & 4.5 & 0\\
                   & M-tracks, \citetalias{sb2013} & 95.1 & 4.8 & 0.1\\
 \hline
\end{tabular}
\tablefoot{Column 1: cosmology and mass range. Column 2: name of the configuration. Column 3: percentage of IMBHs deriving from the collapse of a star ($m_*\leq300\,\msun$) that formed via single and binary stellar evolution. Column 4: percentage of IMBHs forming via VMS collapse. Column 5: percentage of IMBHs forming via BBH merger.}
\label{table:form_chan}
\end{table}

\subsection{Total number of IMBHs}\label{sec:number_all_seeds}

\begin{figure*}[ht]
    \centering
    \includegraphics[width=0.9\textwidth]{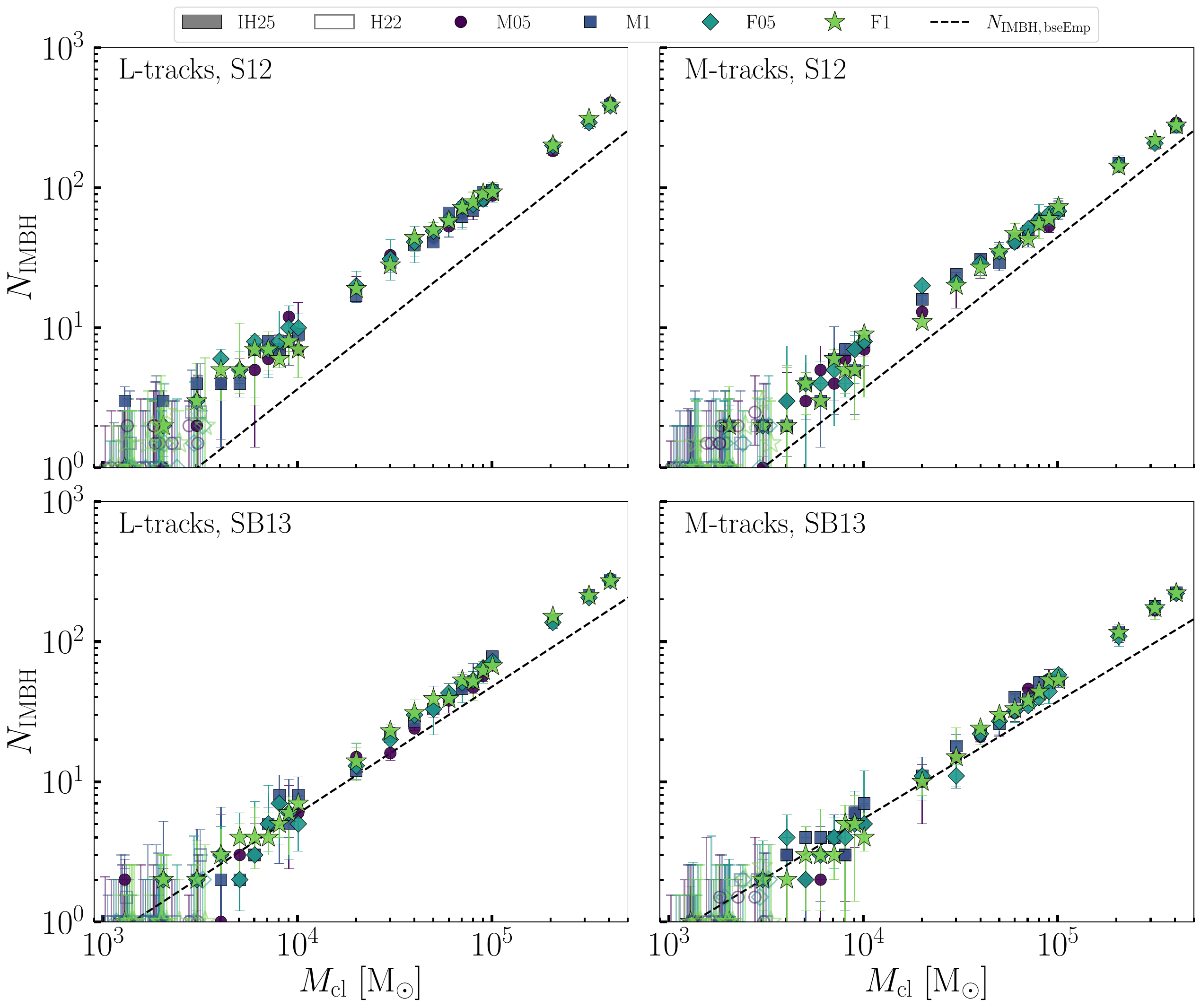}
    \caption{Median of the total number of IMBHs with $90\%$ confidence interval computed over all the repetitions of the simulations performed using \citetalias{hartwig2022} (empty markers) and \citetalias{ishiyama2025} (full markers). The black dashed line represents the number of IMBHs formed in isolation and simulated with \bseemp.}
     \label{fig:num_seeds_cfr}
\end{figure*}

Figure~\ref{fig:num_seeds_cfr} shows the median number of IMBHs formed in clusters after $20\,\rm Myr$, computed across all realizations. For comparison, we also show the number of IMBHs expected from an isolated population of single and binary stars evolved with \bseemp, adopting the same binary fractions as in our dynamical simulations. This allows us to isolate the role of dynamics in the formation of IMBHs. 

In clusters with $M_{\rm cl}\lesssim3000\,\msun$, the number of IMBHs is largely independent of cluster mass across all configurations, for both \citetalias{hartwig2022} and \citetalias{ishiyama2025}. 
We also find that the initial dynamical configuration and density has little impact on the total number of IMBHs, even though it significantly affects the maximum IMBH mass (see Appendix~\ref{app:maximum_seeds}).

The configuration combining L-tracks stars with orbital parameters from \citetalias{sana2012} produces the largest number of IMBHs, with an efficiency $\eta = N_{\rm IMBH}/M_{\rm cl} \sim 10^{-3}\,\msun^{-1}$. Models with M tracks and \citetalias{sana2012}, and models with L tracks and \citetalias{sb2013} result in similar efficiencies ($\eta\sim8\times10^{-4}\,\msun^{-1}$), while M tracks with \citetalias{sb2013} yield the lowest number of IMBHs ($\eta\sim5\times10^{-4}\,\msun^{-1}$).

We find that, due to the effect of dynamical encounters on tight binaries, models with orbital parameters from \citetalias{sana2012} show the largest deviations from the isolated channel, whereas those adopting \citetalias{sb2013} follow it more closely. The setup combining L-tracks stars with and \citetalias{sana2012} yields the most favorable environment for IMBH formation. In models with \citetalias{sb2013}, instead, the number of IMBHs forming in clusters with $M_{\rm cl}\lesssim10^4\,\msun$ can be smaller than the one expected in isolation. In fact, wide primordial binaries tend to break easily in these star clusters, without being able to pair up dynamically. Finally, across all our models, repeated stellar collisions boost the number of IMBHs in clusters with $M_{\rm cl}\gtrsim2\times10^4\,\msun$ compared to the isolated case.

\section{Discussion}\label{sec:discussions}

\subsection{Comparison with previous works}\label{sec:cfr_lit}

In this work, we have investigated the number densities and properties of IMBHs forming in Pop.~III star clusters, assuming a variety of different configurations, and starting from a set of cosmologically-motivated initial conditions. 

By $z\sim19$, we find that IMBHs with masses $m_{\rm IMBH}\sim200\,\msun$ reach number densities of $n_{\rm IMBH}\sim10^{-1}-5\,\rm cMpc^{-3}$. For higher masses, the number density decreases to $n_{\rm IMBH}\sim10^{-2}\,\rm cMpc^{-3}$ at $m_{\rm IMBH}\sim10^3\,\msun$ and $n_{\rm IMBH}\lesssim10^{-3}\,\rm cMpc^{-3}$ for $m_{\rm IMBH}\gtrsim5000\,\msun$. These results are largely insensitive to the adopted stellar evolution models and binary orbital parameter prescriptions, but highly sensitive to the existence of dense massive ($M_{\rm cl}\gtrsim10^4\,\msun$) Pop.~III star clusters.

Previous studies have estimated the number density of light SMBH seeds from Pop.~III stars to be $n\sim1\,\rm cMpc^{-3}$ for $m\sim\mathcal{O}\,(10^2)\,\msun$ at $z\sim20-10$ \citep[e.g.,][]{madau2001, regan2020, regan2024, mehta2026, prole2026}. Our results are broadly consistent with these estimates, but highlight a strong dependence on the clustering properties of Pop.~III stars. In particular, if stars formed preferentially in massive clusters \citepalias{ishiyama2025}, the number density of IMBHs could be enhanced by up to a factor of $\sim5$ compared to the isolated case. Conversely, if Pop.~III stars formed in smaller clusters \citepalias{hartwig2022}, the resulting number densities would be comparable to, or smaller than, those expected for isolated stars.

For more massive seeds, previous works relying on strong Lyman-Werner backgrounds predict significantly lower abundances, with $n\sim10^{-10}-10^{-5}\,\rm cMpc^{-3}$ at $z\sim10$ for $m\gtrsim10^3\,\msun$ \citep[e.g.,][]{regan2024, obrennan2025}. However, alternative formation channels involving rapidly growing halos can produce massive clusters ($\sim9\times10^4\,\msun$) by $z\sim15$, yielding IMBHs of $\sim6000\,\msun$ with number densities $n\sim10^{-5}-10^{-1}\,\rm cMpc^{-3}$ \citep{regan2020, regan2024}. Our simulations show that dense and massive Pop.~III clusters can form IMBHs up to $\sim6200\,\msun$ already by $z\sim19$, with number densities up to $\sim10^{-3}\,\rm cMpc^{-3}$. Consequently, these environments provide a particularly efficient pathway for the formation of high-mass IMBHs, primarily through the direct collapse of VMSs. Importantly, in most cases ($\gtrsim88\%$) these IMBHs remain bound to their parent clusters.

\subsection{Perspectives for growth and SMBH seeding}\label{sec:seeds_growth}

Our results provide new insight into the role of stellar dynamics in shaping the population of light and heavy SMBH seeds. Only a small fraction of the IMBHs we find ($\lesssim10^{-2}\,\rm cMpc^{-3}$) would be required to match the observed number density of SMBHs at high redshift \citep[e.g.,][]{maiolino2024,akins2025,greene2026,umeda2026}, and even today \citep{greene2020}. This implies that even if Pop.~III stars clustered with efficiencies comparable to those in the local Universe \citep[$\sim10\%$,][]{krujissen2012, adamo2015}, the resulting IMBH population could still represent a viable seeding channel.

A key advantage of IMBHs forming via stellar evolution or VMS collapse is that they are typically retained within their host clusters. We recall that only $1-12\%$ IMBHs with typical masses $\lesssim500\,\msun$ are ejected from their host clusters. Because star clusters are quite massive, they experience strong dynamical friction, allowing them and their embedded IMBHs to sink efficiently toward galactic centers \citep[e.g.,][]{antonini2013, smith2018, pfister2019,torniamenti2026}. Once in the galactic nuclei, IMBHs may grow through gas accretion and mergers with other BHs, potentially reaching supermassive scales \citep[e.g.,][]{mapelli2021,natarajan2021,paiella2026}. However, feedback from newly forming stars and from the BH itself may prevent their growth \citep[e.g.,][]{Shi2026}.

The hierarchical assembly of structure further supports IMBH growth. Minihalos hosting star clusters are expected to undergo frequent mergers, which can lead to the formation of more massive and denser clusters. This process could enhance the rate of stellar collisions and BH mergers, further promoting IMBH growth \citep[e.g.,][]{souvaitzis2025}. Alternatively, if one of the merging halos still contains significant gas, this material may be accreted onto the cluster, potentially fueling both additional star formation and further mass growth of the IMBHs. 

In a follow-up study, we plan to investigate the impact of mini DM halo mergers on the evolution of their embedded star clusters using direct $N$-body simulations. In particular, we will explore the role of a time-dependent external potential and assess how such dynamical environments affect cluster structure, survival, and the growth pathways of IMBHs.

\subsection{Caveats}\label{sec:caveats}

Our study relies on initial conditions motivated by the semi-analytical models of \citetalias{hartwig2022} and \citetalias{ishiyama2025}. To account for uncertainties in Pop.~III star formation, we have explored a wide range of stellar evolution models \citep{tanikawa2020}, binary parameter distributions \citepalias{sana2012, sb2013}, and initial cluster configurations, including both monolithic and fractal setups. Our fiducial model adopts a log-flat IMF, with alternative prescriptions discussed in Appendix~\ref{app:imf_impact}.

A key assumption of our work is that the initial mass of the star cluster corresponds to the total Pop.~III stellar mass formed in each mini-DM halo. In \citetalias{ishiyama2025}, however, no multiplicity is assumed and the resulting masses correspond to individual Pop.~III stars. Our approximation is uncertain because fragmentation, particularly in halos with non-zero baryon-DM streaming velocities, may redistribute the stellar mass among multiple stars while also reducing the total stellar mass formed in a halo \citep[e.g.,][]{hirano2018, liu2024b, hirano2026}. Consequently, the cluster masses inferred from \citetalias{ishiyama2025} should be viewed as upper limits, implying that our most massive clusters likely represent an optimistic scenario for SMBH formation.

We adopt half-mass radii of $r_{\rm h}=0.5$ and $1\,\rm pc$, consistent with observations and models of lower-redshift star clusters \citep{marks2012}, and comparable to, or larger than, the half-mass radii predicted by hydrodynamical simulations of Pop.~III star clusters \citep{liu2024b}. Nevertheless, the structural properties of primordial star clusters remain highly uncertain. The efficiency of stellar interactions and collisions in our simulations depends sensitively on the initial cluster density. If typical Pop.~III clusters were less dense, or had masses comparable to (or smaller than) those predicted by \citetalias{hartwig2022}, the resulting IMBH population would likely be similar to, or even less abundant than, that expected from isolated Pop.~III stars. Conversely, if even a modest fraction of minihalos hosted denser and more massive clusters (e.g., with properties intermediate between those predicted by \citetalias{hartwig2022} and \citetalias{ishiyama2025}) a substantial population of both light and heavy SMBH seeds could already be in place by $z\sim19$.

Our $N$-body simulations do not include several physical processes that may affect cluster evolution. In particular, we neglect cluster rotation, which has recently been shown to accelerate early mass segregation and potentially enhance the rate of stellar collisions \citep{kamlah2022, bianchini2026}. We do not model the presence of gas within the clusters, nor do we follow subsequent star formation after the first supernova explosions. Both gas accretion onto BHs and the formation of Population~II stars could significantly impact the growth of IMBHs. However, we expect that our current star clusters do not retain a large fraction of gas because of their low escape velocities ($v_{\rm esc}<50\,\rm km\,s^{-1}$) compared to the typical velocities of stellar winds and SN ejecta. We plan to investigate these effects in future work.  

Finally, the \petar code does not include GW recoil kicks following compact object mergers. As a consequence, we assume that IMBHs can only form from first-generation BBH mergers, and that they are consequently ejected. This assumption is justified by the low escape velocities in our clusters. However, \petar still retains $n$-generation BH remnants, which could affect the dynamics of the cluster.

\section{Conclusions}\label{sec:summary}

We investigated the properties and number densities of IMBHs forming in Pop.~III star clusters at $z\sim20$, using the direct $N$-body code \petar, and cosmologically motivated initial conditions from \citet[][hereafter \citetalias{hartwig2022}]{hartwig2022} and \citet[][hereafter \citetalias{ishiyama2025}]{ishiyama2025}. To account for the large uncertainties in Pop.~III stellar populations, we explored different stellar evolutionary tracks, orbital parameter distributions, and cluster dynamical configurations (Fig.~\ref{fig:init_cond}).

Our main findings can be summarized as follows:
\begin{itemize}[label=\textbullet]
    \item Across all models, and for both sets of cosmological initial conditions, the IMBH number density, $n_{\rm IMBH}$, peaks at $m_{\rm IMBH}\sim200\,\msun$. For IMBH masses $\lesssim500\,\msun$, number densities are comparable to or larger than those expected from isolated Pop.~III stars, reaching $n_{\rm IMBH}\sim0.2-5\,\rm cMpc^{-3}$. In the most massive and dense clusters from \citetalias{ishiyama2025}, IMBHs with $m_{\rm IMBH}\in[1000,6200]\,\msun$ form efficiently already by $z\sim19$, with number densities $n_{\rm IMBH}\sim10^{-4}-10^{-2}\,\rm cMpc^{-3}$ (Fig.~\ref{fig:n_seed}).
    \item The majority ($\gtrsim65\%$) of IMBHs originate from the collapse of stars with $m_*\leq300\,\msun$ forming via single and binary stellar evolution or from stellar collisions (Table~\ref{table:form_chan}). In the more massive and dense clusters of \citetalias{ishiyama2025}, the very massive star (VMS, $m_*>300\,\msun$) channel becomes more important ($\lesssim35\%$), while IMBH formation through BBH mergers remains rare in all simulations ($\lesssim0.1\%$). The most massive IMBHs forming in our simulations are always produced through VMS collapse (Fig.~\ref{fig:imbh_growth_track}). Subsequent stellar collisions (BBH mergers) can further increase IMBH masses in $\lesssim10\%$ ($\lesssim1\%$) of the cases. 
    \item Assuming different stellar tracks (with lower or larger overshooting) and different initial orbital parameter distributions has a crucial impact on the total number of IMBHs formed in Pop.~III stellar clusters (Fig.~\ref{fig:num_seeds_cfr}). The combination of large stellar radii and tight binaries (L-tracks stars with \citetalias{sana2012}) produces the highest IMBH abundances because of more efficient binary interactions. Conversely, models combining compact stars (M tracks) with the wider orbital distributions of \citetalias{sb2013} yield the smallest IMBH populations relative to isolated Pop.~III evolution.
\end{itemize}

Our results indicate that Pop.~III star clusters represent efficient environments for the formation of both light and heavy SMBH seeds, with massive star clusters forming IMBHs up to $\sim6200\,\msun$ already by $z\sim19$. The vast majority of these IMBHs remain bound to the parent clusters ($\gtrsim88\%$), making them also more likely to sink toward galactic centers and grow into SMBHs. This conclusion is largely insensitive to the adopted stellar evolution tracks, orbital parameter distributions, and cluster dynamical configurations. Moreover, it remains valid even if only a fraction ($\sim10\%$) of Pop.~III stars form in sufficiently dense and massive clusters. If Pop.~III star clusters were only as massive as $M_{\rm cl}\sim3000\,\msun$, they could create a population of light seeds comparable or smaller than the one due to isolated Pop.~III stars. In future work, we will investigate how the subsequent growth of DM halos and the hierarchical assembly of star clusters affect the evolution and growth of these dynamical IMBHs.

\begin{acknowledgements}
BM acknowledges INAF (Pleiadi Call 6) for the awarded computational time (project No.~INA24C6B08). BM thanks Lucio Mayer for the very formative discussions had at UZH.

The authors acknowledge financial support from the European Research Council for the ERC Consolidator grant DEMOBLACK, under contract no. 770017 (PI: M. Mapelli) and from the German Excellence Strategy via the Heidelberg Cluster of Excellence (EXC 2181 - 390900948) STRUCTURES. MB and MM acknowledge support from the PRIN grant METE under the contract no. 2020KB33TP. 

MAS acknowledges funding from the European Union’s Horizon 2020 research and innovation program under the Marie Skłodowska-Curie grant agreement No.~101025436 (project GRACE-BH) and from the MERAC Foundation. 

ST acknowledges financial support from the Alexander von Humboldt Foundation for the Humboldt Research Fellowship.

This work was made possible by funding from the French National Research Agency (grant ANR-21-CE31-0026, project MBH\_waves) and from the Centre National d’Etudes Spatiales (MV).

The authors acknowledge support by the state of Baden-W\"urttemberg through bwHPC and the German Research Foundation (DFG) through grants INST 35/1597-1 FUGG and INST 35/1503-1 FUGG.

RSK acknowledges financial support from the ERC via Synergy Grant ``ECOGAL'' (project ID 855130). In addition RSK is grateful for funding from the German Ministry for Economy and Energy (BMWE) in project ``MAINN'' (funding ID 50OO2206), and from DFG and ANR for project ``STARCLUSTERS'' (funding ID KL 1358/22-1). 

SH acknowledges financial support from JSPS KAKENHI Grant Numbers JP21K13960 and JP26K00743.

VL ackowledges computing resources provided by the Ministry of Science, Research and the Arts (MWK) of the State of Baden-W\"{u}rttemberg through bwHPC and the German Science Foundation (DFG) through grants INST 35/1134-1 FUGG and 35/1597-1 FUGG, data storage at SDS@hd funded through grants INST 35/1314-1 FUGG and INST 35/1503-1 FUGG.

TI has been supported by IAAR Research Support Program in Chiba University Japan, and MEXT/JSPS KAKENHI (Grant Number JP26H02062).

The simulations were carried out on two high-performance computing facilities: the \texttt{CINECA Leonardo Booster}, using a partition with 256 NVIDIA A100 GPUs and 64 Intel Xeon Platinum 8358 CPUs (32 cores each), and the \texttt{bwForCluster Helix}, equipped with 4--8 NVIDIA A100 GPUs and two AMD EPYC CPUs (64 cores each).

We obtained our star formation rate densities from \textsc{a-sloth} (\url{https://gitlab.com/thartwig/asloth}) and from the work of \citet{ishiyama2025}. 

This research made use of the Python packages \textsc{NumPy} \citep{harris2020}, \textsc{SciPy} \citep{scipy2020}, \textsc{Pandas} \citep{pandas2020}, and \textsc{Matplotlib} \citep{hunter2007}. The simulations were performed with the direct $N$-body code \petar \citep{wang2020b} (\url{https://github.com/lwang-astro/PeTar}) and the population-synthesis code \bseemp \citep{tanikawa2020}. We gratefully acknowledge Long Wang and Ataru Tanikawa for developing and making these codes publicly available.

\end{acknowledgements}

\bibliographystyle{aa} 
\bibliography{bibliography} 

\begin{appendix}

\section{Impact of the initial mass function}\label{app:imf_impact}

To assess the role of the IMF on the results shown in Fig.~\ref{fig:n_seed}, we compare our fiducial log-flat IMF with a ``super'' top-heavy IMF \citep[\citetalias{sb2013};][]{jaacks2019, liu2020, costa2023}, defined as $\xi(m)\propto m^{-0.17}\,\exp{(-m_{\rm cut}^2/m^2)}$, with $m_{\rm cut}=20\,\msun$. For both IMFs, we consider two values of the maximum stellar mass, $m_{\rm max}=150$ and $300\,\msun$. These simulations were performed with a single realization, and considering the most favorable configuration for IMBH formation, namely cosmological initial conditions from \citetalias{ishiyama2025}, fractal initial conditions with $r_{\rm h}=0.5\,\rm pc$ (F05), and orbital parameters from \citetalias{sana2012}. 

\begin{figure}[H]
    \centering
    \includegraphics[width=0.9\columnwidth]{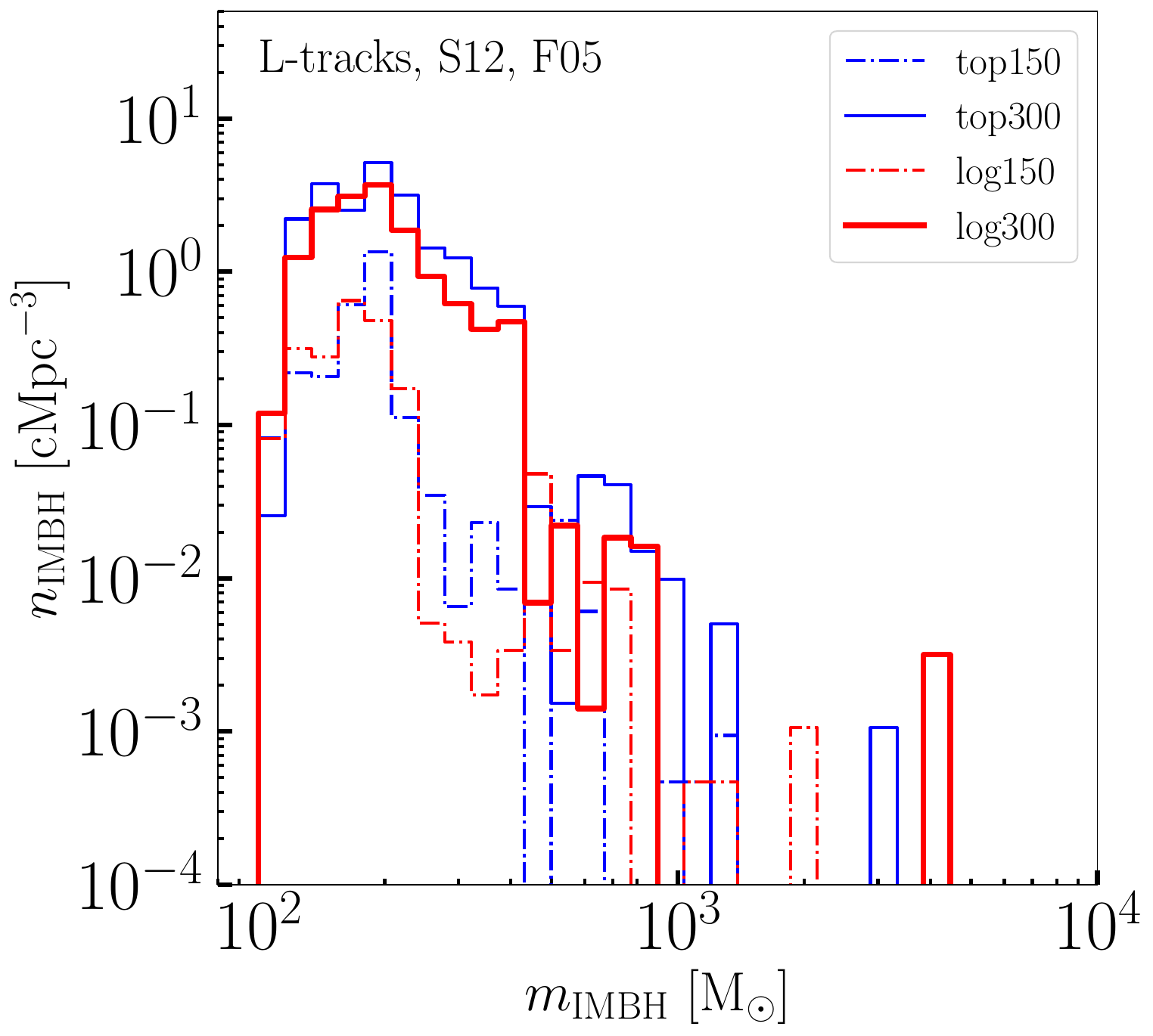}
    \caption{Same as Fig.~\ref{fig:n_seed}, but considering initial conditions from \citetalias{ishiyama2025}, L-tracks, \citetalias{sana2012}, and F05. We show the weighted number density for a log-flat (red) and a top-heavy IMF (blue). We consider $m_{\rm max}=300\,\msun$ (solid lines) and $m_{\rm max}=150\,\msun$ (dot-dashed lines). The thick red line highlights the fiducial model used throughout this work (log-flat, $m_{\rm max}=300\,\msun$). All the simulations were performed only once.}
     \label{fig:num_seed_imf}
\end{figure}

Figure~\ref{fig:num_seed_imf} shows that adopting a more top-heavy IMF with $m_{\rm max}=300\,\msun$ leads to a modest increase in the number density of IMBHs across the entire mass spectrum, relative to the fiducial case. This behavior is expected, as the larger fraction of massive stars enhances the probability of producing massive remnants through the discussed formation channels.

When instead adopting $m_{\rm max}=150\,\msun$, both models retain a peak at $m_{\rm seed}\sim200\,\msun$, but show a significantly different overall trend. In particular, the log-flat IMF produces approximately one order of magnitude fewer IMBHs at the peak, while the top-heavy IMF yields a reduction by a factor of $\sim5$. In addition, both IMFs produce a deficit of objects in the range $\sim200-400\,\msun$, especially in the log-flat case. In fact, when $m_{\rm max}=300\,\msun$, IMBHs form in this range of mass through single stellar collisions. Instead, if $m_{\rm max}=150\,\msun$, that range can be filled only through repeated stellar collision able to produce stars above the PISN gap \citep{mestichelli2026}. 

Finally, both the log-flat and top-heavy IMFs are capable of producing IMBHs with masses $\gtrsim10^3\,\msun$. If $m_{\rm max}=300\,\msun$, number densities reach up to $n_{\rm IMBH}\sim {\rm few}\times10^{-2}\,\rm cMpc^{-3}$. With $m_{\rm max}=150\,\msun$, instead, IMBHs above $10^3\,\msun$ form with number densities up to $n_{\rm IMBH}\sim10^{-3}\,\rm cMpc^{-3}$.

\section{Properties of the most massive IMBHs}

\subsection{H22}\label{app:h22_maxseed}
\begin{figure*}[ht]
    \centering
    \sidecaption
    \includegraphics[width=12cm]{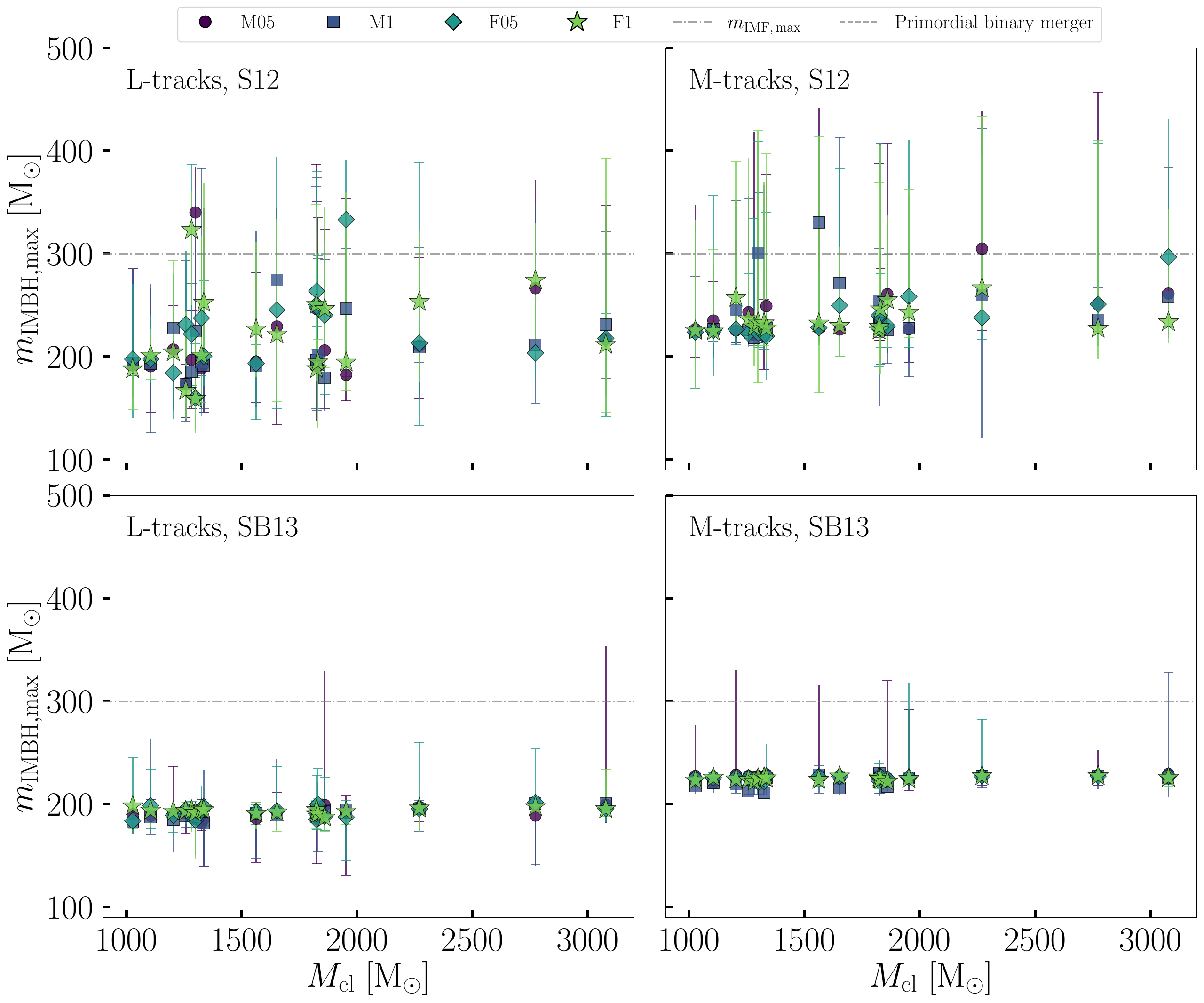}
    \caption{Median of the maximum mass of the IMBHs with $90\%$ confidence interval computed over all the repetitions of the simulations performed using \citetalias{hartwig2022}. The markers represent the different dynamical configurations (M05, M1, F05, F1). The gray dotted-dashed line shows the upper mass limit of the IMF.}
     \label{fig:max_seed_h22}
\end{figure*}

Figure~\ref{fig:max_seed_h22} shows the median of the maximum IMBH masses obtained across all realizations of the simulations adopting the initial conditions from \citetalias{hartwig2022}. We find that, for the relatively small cluster masses predicted by \citetalias{hartwig2022}, the maximum IMBH mass shows little dependence on either the cluster mass or the initial dynamical configuration. In almost all cases, the median maximum IMBH mass is $\sim200\,\msun$. In the case of M-tracks stars, the median is slightly above this value, both because of the mass retention of these stars due to their small radii, and because of the larger PISN gap. When adopting the orbital parameter distributions from \citetalias{sana2012}, the smaller initial semi-major axes increase the probability of interactions between stars in binaries, even for systems evolving on M tracks. As a consequence, two effects occur. First, the PISN mass gap can be partially populated, particularly for stars evolving along the L-tracks. Second, binary interactions can lead to the formation of IMBHs more massive than $300\,\msun$, through stellar mergers and the subsequent collapse of VMSs (see Sec.~\ref{sec:formation_channels}).

\subsection{IH25}\label{app:maximum_seeds}

\begin{figure*}[ht]
    \centering
    \sidecaption
    \includegraphics[width=12cm]{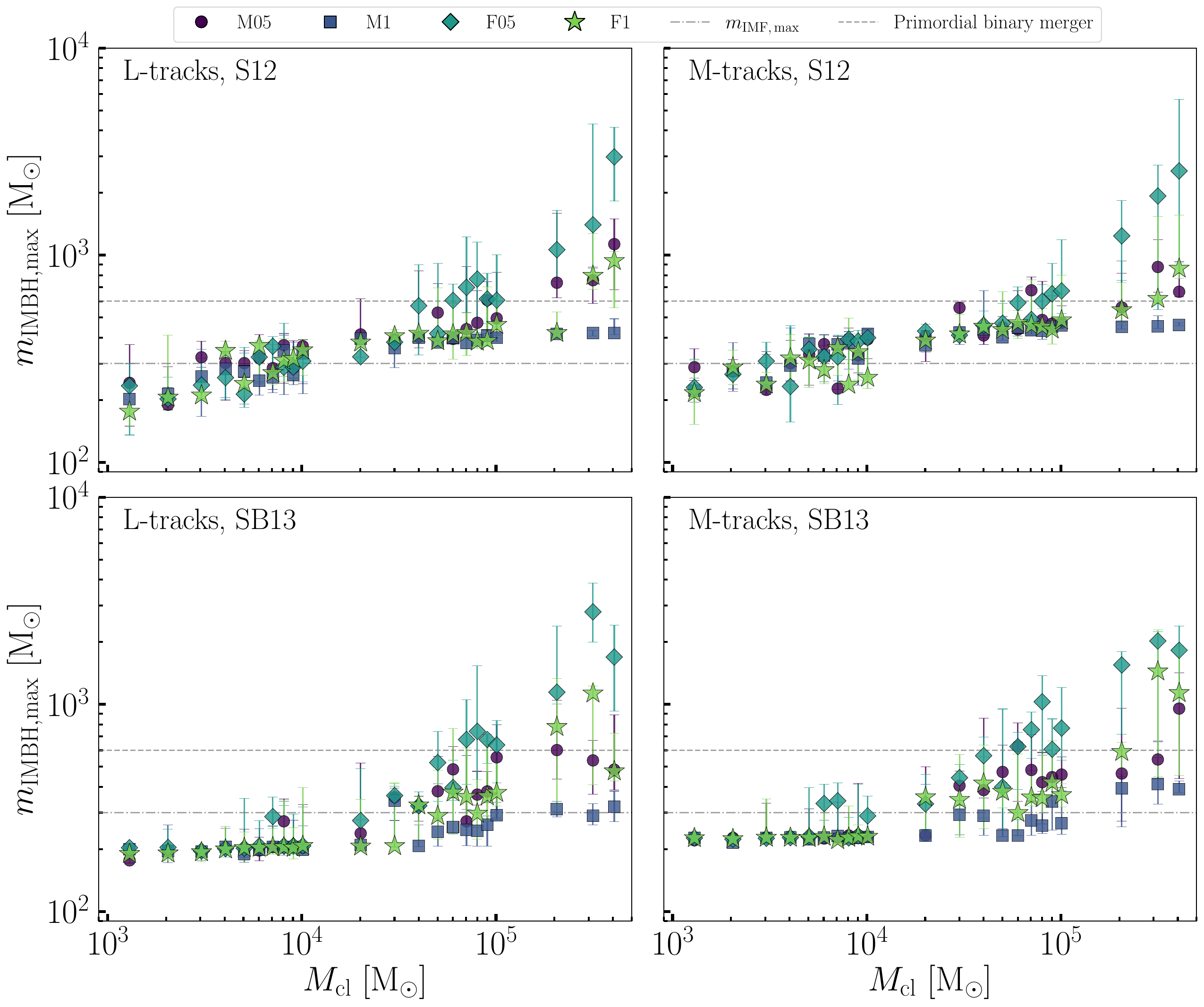}
    \caption{Same as Fig.~\ref{fig:max_seed_h22} but for \citetalias{ishiyama2025}. The gray dotted line represents the maximum mass obtained through a primordial binaty merger.}
     \label{fig:max_seed_ih25}
\end{figure*}

Figure~\ref{fig:max_seed_ih25} shows that, when we consider initial conditions from \citetalias{ishiyama2025}, the median maximum IMBH mass rises with cluster mass $M_{\rm cl}$, from $\sim200\,\msun$ to $\sim5000\,\msun$. The PISN mass gap is most populated in models combining L-tracks and \citetalias{sana2012}, because massive stars on L-tracks reach larger radii and, in tighter binaries, are more likely to undergo envelope stripping during interactions.

At cluster masses comparable to those predicted by \citetalias{hartwig2022}, models based on \citetalias{ishiyama2025} produce IMBHs with similar masses (Fig.~\ref{fig:max_seed_h22}). 

In models adopting orbital parameters from \citetalias{sana2012}, the maximum IMBH mass remains in the range $200-400\,\msun$ up to $M_{\rm cl}\sim10^4\,\msun$. In this regime of cluster masses and densities, IMBHs form through a combination of single and binary stellar evolution channels, and stellar collisions. When instead adopting the wider initial orbital separations from \citetalias{sb2013}, IMBHs form primarily via single stellar evolution and have typical masses of $\sim200\,\msun$. 

Across all configurations, the maximum IMBH mass remains below $\sim600\,\msun$ up to $M_{\rm cl}\sim5\times10^4\,\msun$. At these masses, IMBHs can originate from binary interactions between massive stars or from the collapse of VMSs. We note that, for models adopting \citetalias{sb2013}, the least dense clusters in this mass range produce maximum IMBH masses below $300\,\msun$. 

Finally, for cluster masses $\gtrsim {\rm few}\times10^4\,\msun$, the most massive IMBHs exceed $600\,\msun$ and are typically produced through dynamical processes, such as the collapse of VMSs formed via repeated stellar collisions and, less frequently, through BBH mergers. Above $M_{\rm cl}\sim4\times10^4\,\msun$, models with fractal initial conditions (particularly F05) are the most effective at producing IMBHs with masses exceeding $10^3\,\msun$. Fractal initial conditions can efficiently produce massive IMBHs in large clusters, even when the stellar and binary initial conditions are otherwise unfavorable (see Sec.~\ref{sec:formation_channels}).

 \end{appendix}

\end{document}